\documentclass[prl,amsmath,amssymb,superscriptaddress,twocolumn]{revtex4}
\usepackage{graphicx,color}
\usepackage{bbm}
\newcommand{\kD}{{k_D}}

\newcommand{\br}{{\bf r}}
\newcommand{\bx}{{\bf x}}
\newcommand{\by}{{\bf y}}
\newcommand{\bz}{{\bf z}}
\newcommand{\bk}{{\bf k}}

\newcommand{\bv}{{\bf v}}

\newcommand{\bA}{{\bf A}}

\newcommand{\bG}{{\bf G}}
\newcommand{\bK}{{\bf K}}
\newcommand{\bR}{{\bf R}}
\newcommand{\bdelta}{\delta}

\newcommand{\alphaD}{{\alpha_{D}}}

\DeclareMathAlphabet{\mathpzc}{OT1}{pzc}{m}{it}
\begin{document}
\title{Quantum oscillations in the mixed state of d-wave superconductors}
\author{Ashot Melikyan}
\affiliation{Materials Science Division, Argonne National Laboratory,
Argonne, IL, 60439}
\author{Oskar Vafek}
\affiliation{National High Magnetic Field Laboratory and Department
of Physics,\\ Florida State University, Tallahassee, Florida 32306,
USA}
\date{\today}
\begin{abstract}
We show that the low-energy density of quasiparticle states in the
mixed state of ultra-clean $d-$wave superconductors is characterized
by pronounced quantum oscillations in the regime where the cyclotron
frequency $\hbar\omega_c\ll \Delta_0$, the $d-$wave pairing
gap. Such oscillations as a function of magnetic field $B$ are argued to
be due to the internodal scattering of the $d$-wave quasiparticles
near wavevectors $(\pm \kD,\pm\kD)$ by the vortex lattice as well as
their Zeeman coupling. The
periodicity of the oscillations is set by the condition
$\kD \sqrt{hc/(eB)} \equiv  \kD' \sqrt{hc/(eB')}\pmod {2\pi}$. We find that there is additional
structure within each
period which grows in complexity as the Dirac node anisotropy
increases.
\end{abstract}
\maketitle

The behavior of $d$-wave superconductors in magnetic field has
recently come into sharp focus due to the experimental observation
of the oscillations in the longitudinal and Hall electrical
transport at extremely high magnetic fields for very clean
underdoped $\rm YBa_2Cu_3O_{6.5}$ \cite{Doiron2007}
and $\rm YBa_2Cu_4O_8$ \cite{Yelland2007, Bangura2007}. At temperatures
about $2$
K and for magnetic fields in excess of about $30$ T, the Hall
resistance is finite, {\em negative} (i.e. electron like), and
oscillates as a function of magnetic field, with approximately four
pronounced peaks between $\sim 45$ T and $60$ T. Below about $30$
T, the system is in the mixed superconducting state, and the
electrical Hall conductivity vanishes. The current interpretation of
these oscillations involves charge or spin order induced
reconstruction of the Fermi surface with electron pockets
(\cite{Doiron2007,Bangura2007,ChenZhang2007,MillisNorman2007,Chakravarty2007}),
which result from the appropriate tuning of the strength of the
charge/spin potentials in the particle-hole channel
\cite{ChenZhang2007,MillisNorman2007,Chakravarty2007}.

As the estimated mean-field $H_{c2}$ for the YBCO samples in
question significantly exceeds $60$ T\cite{Ong2002}, it is at present
unclear why one should expect the superconducting order parameter
amplitude to collapse at the fields applied in Refs.
\cite{Doiron2007,Bangura2007,Yelland2007}. Instead, the local
superconducting correlations should persist, and rather the system
is expected to be in the vortex liquid state
\cite{HuseFisherFisher1992,LuLi2007}. Additionally, the sign
reversal of the Hall effect, from positive to negative, with
decreasing temperature near T$_c$ in optimally doped YBCO has been
observed and interpreted earlier as resulting from the flux
flow\cite{Galffy1988,Lobb1990,Ong1991}. The structure factor of such
liquid state is expected to closely resemble that of a vortex solid.
In this context, the natural question is the existence, the
periodicity, and the physical nature of such magnetic field induced
oscillations in the vortex state.

In this work we therefore examine the properties of the $d-$wave
quasiparticles (qps) in the vortex solid state at intermediate
magnetic fields, i.e. we examine the particle-particle channel. For
the square vortex lattice observed in the small angle neutron
scattering\cite{Brown2004}, we find that the combined effect of the
orbital and spin coupling of the qps to the magnetic field does
induce oscillations in the low energy density of states, $N(E)$, and
that the nature of these oscillations depends on the Dirac cone
anisotropy $\alphaD$. In the physically relevant regime where the
$d$-wave gap $\Delta_0\gg \hbar\omega_c$ (the cyclotron
frequency $\omega_c=eB/mc$), the density of states $N(E)$ is an
oscillatory function of $\kD\ell$, where $\kD$ is the $k-$space
half-distance between the nearest nodes (see Fig. 1) and $\ell$ is
the magnetic length $\ell=\sqrt{hc/Be}$.
For $\alphaD=1$, the period of the oscillations
corresponds to $\delta(\kD\ell)=2\pi$, but with increasing
anisotropy the number of oscillations {\it within} the period
increases. In the absence of Zeeman coupling $N(E)$ is exactly
periodic in the scaling regime and the spectrum oscillates between
fully gapped and nodal. Upon inclusion of the Zeeman coupling, which
cannot be neglected for fields of
Ref. \cite{Doiron2007,Bangura2007,Yelland2007}, the oscillatory
behavior persists, but the scaling is only approximate. For
$\alphaD=1$ and for the fields in the excess of $30T$, the combined
effect of the orbital and Zeeman coupling leads to a sequence of
transitions and the system oscillates between a thermal metal with
finite $N(0)$ and a thermal insulator with vanishing $N(0)$. A
similar pattern of oscillating behavior holds for large $\alphaD$,
albeit with effectively increased frequency and significantly
decreased gap. The envelope of the $N(0)$ oscillations follows an
approximately $\ell^{-1}$ behavior. These findings should be
contrasted with the oscillations found in the regime of extreme high
fields Ref. \cite{YasuiKita1999} where the number of Landau levels
below the Fermi energy range from $1$ to $7$.

\begin{figure}
\includegraphics[width=0.95\columnwidth,clip]{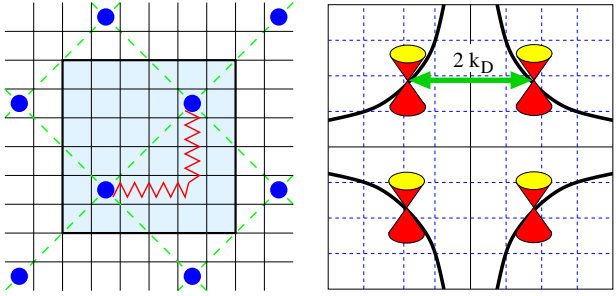}
\caption{\label{Fig:1} Left: Magnetic unit cell $\ell\times\ell$ of
a tight-binging lattice (solid lines) containing two vortices joined
by a branch-cut shown for $\ell=6a$. Right: Fermi surface (black
solid lines) of cuprate superconductors with four nodal points at
$\bk_F^{(j)}=(\pm \kD,\pm \kD )$ in a Brillouin zone of size
$2\pi/a$. In the presence of the vortex lattice, the Brillouin zone
is reduced to $2\pi/\ell$ (spacing of the blue grid). The properties
of the low energy states depend on the commensuration between the
inter-nodal distance $2k_N$ and the reduced reciprocal lattice
vector $2\pi/\ell$.}
\end{figure}
\begin{figure*}
\includegraphics[width=1\textwidth,clip]{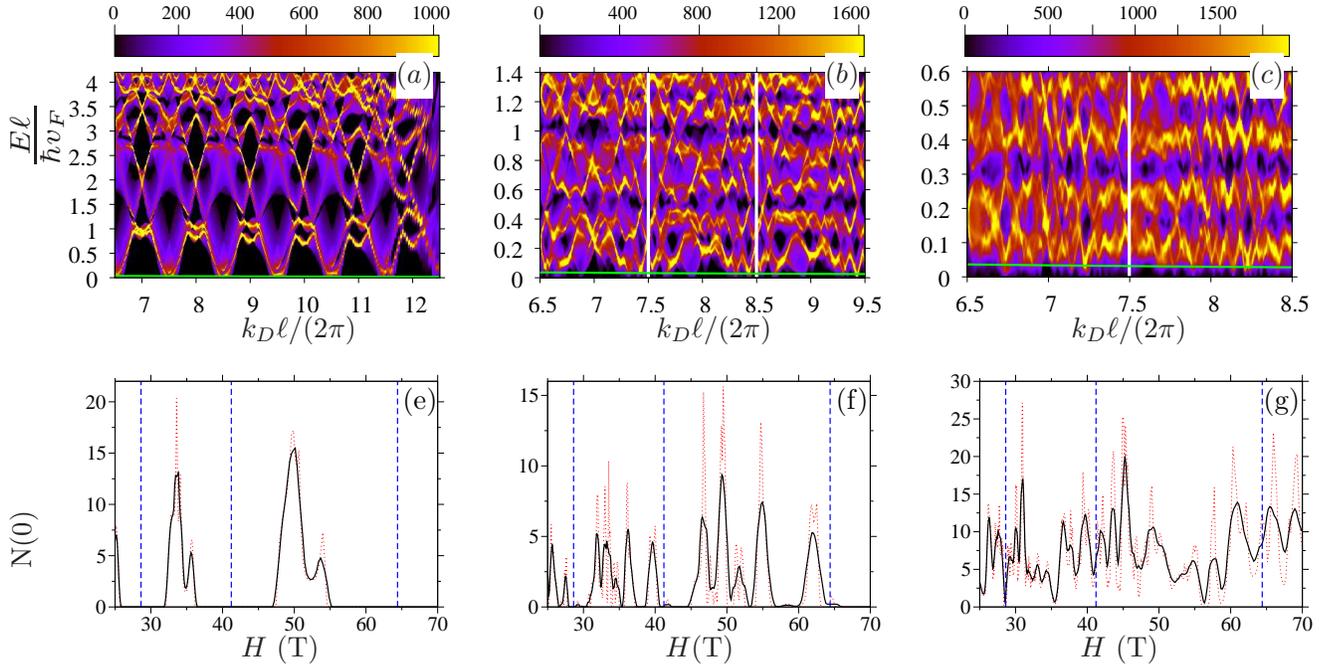}
\caption{\label{Fig:2} (a-c): Oscillations of the qp density of
states $N(E)$ rescaled as $\ell v_F N(\ell E/ (\hbar v_F))$ are shown as a
color map for $\alphaD=1$ (a), $\alphaD=4$ (b), and $\alphaD=7$ (c).
The green line represents the shift of the qp zero energy due to the
Zeeman term for $t=380$ meV, $a=3.8$ \AA, $g=2$, $\kD=0.38\pi/a$, and
$\ell=26a$.
(e-f): Zero energy density of states of (\ref{bdg0}) as a function
of magnetic field for $\alpha_{\Delta} = 1$ (e), $4$ (f), $7$ (g) deduced
from the figures of the top row and the generalized scaling form of
eigenvalues (\ref{simonleescalingNew}) is shown in red dashed line.
The solid black line corresponds to an effective broadening due to
the inter-layer coupling $2t_z\cos(k_z c)$ with $t_z = 0.05t$. }.
\end{figure*}
The starting point is the lattice Bogoliubov-de~Gennes (BdG)
eigenequation\cite{wangmacdonald,vmftvmt}
\begin{eqnarray}
\mathcal{\hat{H}}_0\psi_{\br}=E\psi_{\br}\textcolor{blue}.
\end{eqnarray}
where the Hamiltonian $\mathcal{\hat{H}}_0$ acts on the two
component Nambu spinor $\psi_{\br}=[\mathpzc{u}_{\br},
\mathpzc{v}_{\br}]^T$ and has the following explicit form
\begin{equation}\label{bdg0}
\mathcal{\hat{H}}_0= \left(\begin{array}{cc}
\mathcal{\hat{E}}_{\br}-\mu+\frac{g}{2}\mu_B B & \hat{\Delta}_{\br} \\
\hat{\Delta}^{\ast}_{\br} &
-\mathcal{\hat{E}^{\ast}}_{\br}+\mu+\frac{g}{2}\mu_B B
\end{array}\right).
\end{equation}
Both $\mathcal{\hat{E}}_{\br}$ and $\hat{\Delta}_{\br}$ are defined
through their action on a lattice function $f_{\br}$ as
\begin{eqnarray}
\mathcal{\hat{E}}_{\br}f_{\br}&=&-t\sum_{\bdelta=\pm\hat{\bx},\pm\hat{\by}}
e^{-i\bA_{\br\br+\delta}} f_{\br+\delta},\\
\hat{\Delta}_{\br}f_{\br}&=&\Delta_{0}\sum_{\bdelta=\pm\hat{\bx},\pm\hat{\by}}
e^{i\theta_{\br \br+\bdelta}}\eta_{\bdelta}f_{\br+\delta}.
\end{eqnarray}
In the symmetric gauge, the magnetic flux $\Phi$ through an
elementary plaquette enters the Peierls factor via
$\bA_{\br\br+\hat{x}}=-\pi y \Phi/\phi_0$, $\bA_{\br\br+\hat{y}}=\pi
x \Phi/\phi_0$; the electronic flux quantum is $\phi_0=hc/e$. The
$d$-wave symmetry is encoded in $\eta_{\bdelta}=+(-)$ if $\bdelta
\parallel \hat{\bx}(\hat{\by})$.
Importantly, $\theta_{\br\br'}$ winds by $2\pi$ around each of the
magnetic field induced vortices. The initial Ansatz\cite{vm} for the
pair phases is
\begin{eqnarray}\label{phase}
e^{i\theta_{\br\br'}}\equiv(e^{i\phi_{\br}}+e^{i\phi_{\br'}})/|e^{i\phi_{\br}}+e^{i\phi_{\br'}}|,
\end{eqnarray}
where $\nabla\times\nabla\phi(\br)=2\pi
\hat{\bz}\sum_{i}\delta(\br-\br_i)$ and
$\nabla\cdot\nabla\phi(\br)=0$ where $\br_i$ denotes the vortex
positions. Although the phase of the order parameter does depend on
the precise form of the self-consistency condition, the deviations
from the adopted form are weak, and moreover, both the symmetry of
the phase and its singular part are fixed unambiguously by the
vortex lattice.

Connecting pairs of vortices by branch cuts\cite{vm}, we can define
the singular gauge transformation\cite{ft,vm}
$\mathcal{U}=e^{\frac{i}{2}\sigma_3\phi_{\br}}$ where the Pauli
sigma matrices act on the Nambu spinors. The transformed Hamiltonian
$\mathcal{H}(\bk)=e^{-i\bk\cdot\br}\mathcal{U}^{-1}\;\mathcal{\hat{H}}_0\;\mathcal{U}e^{i\bk\cdot\br}$
becomes
\begin{equation}\label{H0singgauge}
\mathcal{H}(\bk)=\sigma_3\left(
\tilde{\mathcal{E}}_{\br}(\bk)-\mu\right)+\sigma_1\tilde{\Delta}_{\br}(\bk)+\frac{g}{2}\mu_BB,
\end{equation}
where the transformed lattice operators satisfy
\begin{eqnarray}
\mathcal{\tilde{E}}_{\br}(\bk)\psi_{\br}&=&-t\!\!\!\sum_{\bdelta=\pm\hat{\bx},\pm\hat{\by}}
z_{2,\br\br+\delta}\times e^{i\sigma_3 V_{\br\br+\delta}}
e^{i\bk\cdot\delta}\psi_{\br+\delta}\\
\tilde{\Delta}_{\br}(\bk)\psi_{\br}&=&\Delta_{0}\!\!\!\sum_{\bdelta=\pm\hat{\bx},\pm\hat{\by}}
z_{2,\br\br+\delta}\times
\eta_{\bdelta}e^{i\bk\cdot\delta}\psi_{\br+\delta}.
\end{eqnarray}
The physical superfluid velocity enters via the factor
\begin{eqnarray}
e^{i
V_{\br\br'}}
=e^{-i\bA_{\br\br'}}(1+e^{i(\phi_{\br'}-\phi_{\br})})/|1+e^{i(\phi_{\br'}-\phi_{\br})}|
\end{eqnarray}
and represents the lattice analog of the semiclassical (Doppler)
effect\cite{volovik}. The Z$_2$ field $z_{2,\br\br'}=1$ on each bond
except the ones crossing the branch cut where $z_{2,\br\br'}=-1$.

In the magnetic
fields of interest the vortices of YBCO form a square lattice with
primitive vectors oriented along the $d$-wave nodes\cite{Brown2004},
and therefore the transformed Hamiltonian (\ref{H0singgauge}) is
invariant under discrete translations by the primitive vectors
$\bR_1=\ell \hat{x}$ and $\bR_2=\ell\hat{y}$ defining the magnetic
unit cell, reflecting the periodicity of $V_{\br\br'}$ and the
periodic choice of the branch cuts. Consequently, it can be
diagonalized in the Bloch basis. By the Bloch condition
$\mathcal{H}(\bk)$ acts on the periodic functions, and the crystal
wavevector $\bk$ varies continuously within the 1$^{st}$ Brillouin
zone defined by the primitive reciprocal lattice vectors $\bK_1=2\pi
\frac{\hat{x}}{\ell}$, $\bK_2=2\pi \frac{\hat{y}}{\ell}$. Note that,
since the unitary transformation $\mathcal{U}$ is time independent,
the Hamiltonians (\ref{bdg0}) and (\ref{H0singgauge}) have the same
thermodynamic and tunneling density of states.

The above claims follow from our detailed numerical study of the
tight-binding Bogoliubov-de~Gennes Hamiltonian (\ref{H0singgauge}).
Since the Zeeman term is just a simple overall shift of the
qp energies, let us first neglect it. The results of the
numerical diagonalization of (\ref{H0singgauge}) can be summarized
by the following scaling form of the qp eigenenergies:
\begin{equation}
\label{simonleescalingNew} E_{n}(\bk) = \frac{\hbar v_F}{\ell}
\mathcal{F}_{n}\left(\bk \ell,\alphaD, \kD\ell\right).
\end{equation}
Here $v_F$ and $v_{\Delta}$ are the Fermi and gap velocities of the
$d-$wave qps near the nodes $(\pm \kD,\pm\kD)$ (in our model
(\ref{bdg0}) $v_{F} =2\sqrt{2} a t\hbar^{-1} \sin(\kD a)$ and $v_{\Delta} =
v_F\Delta_0/t$);
$\mathcal{F}_{n}$ is a dimensionless scaling function, which {\it
differs} from the Simon-Lee scaling function in the following
important aspect: {\it it depends on $\kD\ell$}. Specifically, in
the scaling limit of $\kD\ell\rightarrow \infty$, the function
$\mathcal{F}_n$ does not approach a uniform value. This is amply
illustrated in the Fig.(\ref{Fig:2}), where we used a color density
plot to represent the density of states per area rescaled as $v_F
\ell N(E)$ for different values of $\kD\ell/(2\pi)$ and $E\ell/(\hbar v_F)$.

As best seen for $\alphaD=1$, near $E=0$ the spectrum is gapped
although for $\kD\ell\approx \pi(2n+1)$ the gap is very small,
vanishing only at discrete set of points as expected for the unitary
class\cite{vmftvmt}. Moreover, the gap scales as $\ell^{-1}$ and
oscillates as a function of $\kD\ell$.
Physically, the oscillations are due to the
strong internodal scattering which arises from the commensurability
of the $d$-wave nodes and the vortex lattice.
As seen from the three
different panels, the pattern of the oscillations depends on the
anisotropy $\alphaD$. With increasing
$\alphaD$, the DOS acquires ever richer structure with multiple
minima and maxima -- for presentation purposes we show only three
periods (separated by white vertical lines) for $\alphaD=4$ and two
periods for $\alphaD=7$. Additionally, as $\alphaD$ increases, the
scaling limit $\Delta_0\gg \hbar\omega_c$ is reached for smaller
magnetic fields (larger $\ell$) and the approximate low energy
scaling (\ref{simonleescalingNew}) holds for magnetic length
$\ell/a\gg 2\pi\sqrt{\alphaD}$.

The magnetic fields of experimental interest ($45-70$ T) correspond
to $\ell \in (20a, 25a)$, which is the regime where the Zeeman term
cannot be neglected. Nevertheless, it is easy to take it into
account, since it corresponds to a simple shift in zero of the qp
energy, represented by the green lines in each of the color DOS maps
in Fig. \ref{Fig:2}. The DOS along the Zeeman "slice" is shown in
Figs. \ref{Fig:2}(e-f). For anisotropy $\alphaD=1$, the spectrum
exhibits pronounced oscillations as a function of magnetic field (or
$\kD$), changing from gapped (thermal insulator) with activated
temperature dependence of the specific heat to gapless (thermal
metal) with $T-$linear specific heat. This $\delta\ell = 2\pi/\kD$
periodic sequence of transitions between thermal metal and thermal
insulator is significantly richer for $\alphaD>1$. Within each such
``primitive'' period, a complex structure develops, and the number
of minima and maxima depends on $\alphaD$. For $\alphaD=7$ and
$\kD=0.38\pi/a$, which are approximately\cite{sutherland} the
physical values expected for the YBCO samples of
Ref. \cite{Doiron2007}, there are four "oscillations" in the field
range $\sim 45-60T$. Remarkably, this is precisely the
experimentally observed "periodicity"\cite{Doiron2007}.

Note that the effect described here occurs deep in vortex state with
a fully developed superconducting amplitude and does not rely on
semiclassical orbits around electron pockets. Rather, as we now
argue, it is a consequence of the commensuration involving the
vortex lattice and the internodal separation (see Fig. \ref{Fig:1}).
In the scaling regime of interest, $\Delta_0\gg \hbar\omega_c$, the
appropriate starting point should be the linearized
approximation\cite{SimonLee1997, ft} of BdG Hamiltonian
(\ref{H0singgauge}). In this approach the low-energy qps with
$E\ll\Delta_0$ are independent and behave as massless Dirac fermions
interacting with vortices via the Doppler shift originating from the
superflow $\bv(\br)=\frac{\hbar}{2}\nabla\phi-\frac{e}{c}\bA$.
Additionally, the wavefunctions contain branch-cuts connecting
vortices pairwise\cite{SimonLee1997,ft,vmftvmt,mtTB,vm, marinelli, vishwanath}. The effective
Hamiltonian\cite{SimonLee1997, ft, marinelli, vishwanath} is then
\begin{equation}
\label{Hlin} \mathcal{H}_{lin}/v_F = \frac{p_x+p_y}{\sqrt{2}}
\sigma_3+ \alphaD \frac{p_y-p_x}{\sqrt{2}}\sigma_1
+m\frac{v_x+v_y}{\sqrt{2}}.
\end{equation}
In this approximation,
$\mathcal{H}_{lin}$ exhibits the Simon-Lee\cite{SimonLee1997}
scaling $\mathcal{H}_{lin} (\br,v_F,v_{\Delta}) = \frac{\hbar v_F}{\ell}
\mathcal{H}_{lin} (\br/\ell, \alphaD,1)$ and the qp spectrum has a
scaling form $E_{n}(\bk) = \frac{\hbar v_F}{\ell}
\mathcal{F}^{(lin)}_{n}(\bk \ell,\alphaD)$.

Now, we consider the corrections due to the terms left out during
linearization. Provided that $\br$ is not in the immediate vicinity
of a vortex core ($r\gtrsim \xi$), the leading non-linear
corrections, such as $m\bv^2(\br)/2$, are typically omitted on the
basis that they are smaller than the terms retained in (\ref{Hlin})
by a factor of $(\kD\ell)^{-1}$ which is small for typical fields of
interest. However, such estimate is incorrect. The effect of the non-linear
terms is amplified\cite{mtTB} by an anomalously large (low energy)
qp wavefunctions, growing as $r^{-1/2}$ near vortex
locations\cite{melnikov,vmftvmt}. This divergence is eventually cut
off only at $r\approx \xi$. A typical ``perturbation'', such as
$m\bv^2(\br)$, results in the following matrix element between two
eigenstates of $\mathcal{H}_{lin}$ at nodes $j$ and $j'$: \(
I=\langle \Psi^{(j)}_{n\bk} | m\bv^2(\br)
|\Psi^{(j')}_{n'\bk'}\rangle \). Due to Bloch symmetry of the
wavefunctions $\Psi$, the Bloch momenta $\bk_F^{(j)}+\bk$ and
$\bk_F^{(j')}+\bk'$ must differ by a reciprocal lattice vector
$\bG$. Importantly, $|I| \propto \frac{\hbar v_F}{\ell} \frac1{\kD\xi}[C_1
+ C_2 \cos(\bR\cdot \bG)]$, {\it which is of the same order as the
terms retained in} (\ref{Hlin}). Here $\bR$ is the primitive vortex
lattice vector and the vortex core size $\xi$ serves as the
small-distance cut-off of the otherwise divergent
integrals\cite{mtTB}. Coefficients $C_{1,2}$, in general depend on
$\bk$ and $\bk'$, but not on $\bk^{(j)}_F$ or $\bk^{(j')}_F$.

Thus, due to the $r^{-1/2}$-behavior of the low-energy wavefunctions
near vortex cores, the matrix elements $I$, and the resulting
corrections to the energies, scale with magnetic length precisely as
the energies of (\ref{Hlin}), namely as $\sim \ell^{-1}$. The
relative magnitude of the corrections due to the non-linearities
compared to (\ref{Hlin}) is determined by the magnitude of the
parameter $(\kD\xi)^{-1}$ rather than $(\kD\ell)^{-1}$. In cuprate
superconductors the former is typically of $\mathcal{O}(1)$ since
$\xi$ is of the order of a few lattice spacings. Thus, if the
parameters of the BdG Hamiltonian (\ref{H0singgauge}) such $\mu$ or
$\ell$ are varied, the resulting qp spectrum evolves in a manner not
captured by (\ref{Hlin}).

Based on the above argument, for the square vortex lattice under
consideration we expect the spectra for nodal momentum $\kD$
($\kD'$) and magnetic length $\ell$ ($\ell'$) to be similar when
\begin{equation}
\kD\ell \equiv \kD'\ell'\pmod{2\pi}.
\end{equation}
Therefore, the Simon-Lee scaling should be generalized to Eq.
(\ref{simonleescalingNew}). In the regime where the scaling
(\ref{simonleescalingNew}) holds the oscillatory part of the
dispersion is fully determined by the product
$\kD\ell$, and therefore the dependence of the spectrum on the magnetic
field can
be determined from the changes in $\kD$\cite{mtTB}. Finally, for $\mu=0$ (or
equivalently $\kD=\pi/(2a)$), the characteristic oscillations found
in this work can be proven rigorously without appealing to the
perturbation theory \cite{vm}. In this case, when ignoring the
Zeeman coupling, the qp spectrum is gapless when $\kD\ell/(2\pi)$ is
a half-integer and gapped when it is an integer.

Since the effect described here is due to the interference effects
between the nodal $d-$wave qps and the ordered vortex positions, it
should be
observable\cite{mvdenver} as long as the thermal length $L_T= \hbar
v_F/(k_B T)$ exceeds
$\ell$. We expect
these oscillations to persist in a vortex liquid state with
strong, albeit only short range, positional order, provided that the vortex
positional correlation length $\xi_p>>L_T>>\ell$.
The true test of these predictions, however, would be an observation of
the high
field quantum oscillations in the low temperature specific heat in
the superconducting state with nearly perfect vortex lattice. The
two parameters which control the pattern of the oscillations are
$\kD$ and $\alphaD$, both of which depend on doping and can be
determined independently.

We thank Z. Te\v{s}anovi\'{c} for useful discussions. A. M. would also
like to thank M. Norman for helpful discussions and a critical reading of the manuscript.

A. M. was supported by the \mbox{U. S.} Dept. of Energy, Office of
Science, under Contract No. DE-AC02-06CH11357. O. V was supported in
part by the NSF grant DMR-00-84173.

\end{document}